\documentclass[prl,twocolumn,showpacs,amsmath,amssymb]{revtex4} 

\usepackage{graphicx}

\begin{document}
\title{Weak localisation in bilayer graphene }
\author{R.V.Gorbachev}
\author{F.V.Tikhonenko}
\author{A.S.Mayorov}
\author{D.W.Horsell}
\author{A.K.Savchenko}
\affiliation{School of Physics, University of Exeter, Stocker
Road, Exeter, EX4 4QL, U.K.}

\begin{abstract}
We have performed the first experimental investigation of quantum
interference corrections to the conductivity of a bilayer graphene
structure. A negative magnetoresistance -- a signature of weak
localisation -- is observed at different carrier densities,
including the electro-neutrality region.  It is very different,
however, from the weak localisation in conventional two-dimensional
systems. We show that it is controlled not only by the dephasing
time, but also by different elastic processes that break the
effective time-reversal symmetry and provide intervalley scattering.
\end{abstract}

\pacs{73.23.-b, 72.15.Rn, 73.43.Qt}

\maketitle

The discovery that a single layer of carbon atoms (graphene
\cite{NovoselovS}) can be manufactured in such a way that the
carrier density is controlled by a gate voltage
\cite{NovoselovS,ZhangPRL} became a breakthrough in creating a new,
high-quality two-dimensional (2D) system. It has remarkable
differences from conventional 2D systems: a linear energy spectrum
and chirality of charge carriers.  These initiated much interest
(mainly theoretical) in exploring graphene's properties. One of them
is the manifestation of well known \cite{WL} quantum interference
corrections to the conductivity.

Interference of electron waves that are scattered by disorder and
form closed trajectories usually decreases the conductivity,
resulting in the so-called weak localisation (WL) of electrons.
Perpendicular magnetic field introduces a phase difference for
interfering carriers and destroys the WL, making measurements of the
(positive) magnetoconductivity a sensitive tool of quantum
interference. Carriers in graphene are chiral, that is, they have an
additional quantum number, pseudospin, arising from the fact that
their wavefunctions are composed of the contributions from two
sublattices. If pseudospin is conserved, the backscattering of
carriers is forbidden and weak antilocalisation rather than WL will occur
\cite{WAL,Morpurgo,McCann}, with a positive correction to
conductance and a negative magnetoconductance. Furthermore, the
quantum interference in graphene will be sensitive not only to
inelastic scattering of carriers, which breaks the phase of the
wavefunction, but also to a number of elastic scattering mechanisms,
especially atomically sharp and topological defects which affect the
carrier chirality \cite{Morpurgo,McCann}. The energy spectrum
``warping'' of the carriers is also of major importance for WL in
graphene \cite{McCann}. The first experiments on WL in single-layer
graphene \cite{Morozov,Wu,Heersche} have indeed shown its high
sensitivity to the details of scattering processes, even full
suppression of the interference effect \cite{Morozov}.

In this work we study WL in a bilayer graphene structure. It has
already been understood that adding a second layer of graphene
changes the system dramatically: the linear dispersion relation
becomes parabolic \cite{Band,McCann06Koshino}. In spite of this,
bilayer graphene is expected to be very different from conventional
2D systems with parabolic spectrum, as well as from single-layer
graphene \cite{McCann06Koshino,Katsnelson,NilssonSnyman,Kechedzhi}
(as measurements of the quantum Hall effect
\cite{NovoselovNZhangN,NovoselovP} have already shown). The carriers
remain chiral, but the Berry phase of the wavefunction (a measure of
chirality) in a bilayer is $2\pi$ and not $\pi$ as in a single layer
(which means that the pseudospin turns twice as quickly in the plane
than the momentum, while in a single-layer it is aligned with the
momentum). As a result, there will be no suppression of
backscattering, and the quantum correction in a bilayer will have
the sign of conventional WL \cite{Morpurgo,Kechedzhi}. Its
magnitude, however, will still be very sensitive to different
elastic processes: WL in each of the two graphene valleys can be
totally suppressed by topological defects \cite{Morpurgo} and
warping of the energy spectrum (the latter is expected to be
stronger in a bilayer than in a single layer \cite{Kechedzhi}). At
the same time, intervalley scattering can partially restore WL.

Our measurements of the positive magnetoconductance of a
gate-voltage controlled bilayer structure have shown the existence
of WL at all studied carrier densities. The detailed analysis of the
results is performed using the theory \cite{Kechedzhi}, which has
allowed us to determine three physical quantities responsible for
the manifestation of WL: the phase breaking time $\tau_{\phi}$ and
two elastic times -- the intervalley scattering time $\tau_{i}$ and
an effective time-reversal symmetry breaking time $\tau_w$.

The sample of rectangular shape (1.5 $\mu$m $\times 1.8$ $\mu$m) is
fabricated by the method of mechanical exfoliation of (HOPG)
graphite developed in \cite{NovoselovS}. The bilayer flake is
deposited on a $n^{+}$Si substrate separated by a 300 nm SiO$_2$
layer. Using e-beam lithography, two Au/Cr contacts were made at
opposite edges of the flake to allow resistance measurements from 40
mK to 4.2 K. The carrier density up to $1.5\times10^{12}$ cm$^{-2}$
is varied by the gate voltage between the flake and $n^{+}$Si (in
accordance with the relation $dn/dV_g=7.3\times10^{10}$
cm$^{-2}$V$^{-1}$ determined by the thickness of SiO$_2$).

\begin{figure}[htb]{}
\includegraphics[width=.98\columnwidth]{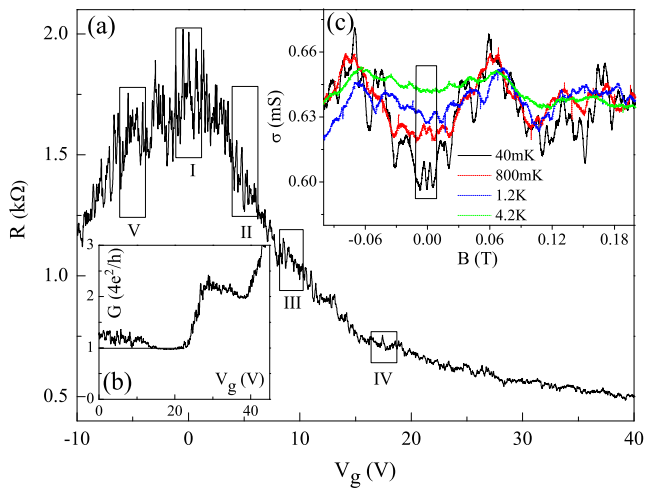}
\caption{(color online) (a) Dependence of the resistance of the bilayer flake on
gate voltage at 40 mK. (b) Conductance as a function of $V_g$ at $B$
= 14.3 T. (c) Conductivity fluctuations as a function of magnetic
field for different temperatures (region II). The rectangle defines
the range of fields used for WL measurements.}\label{fig:G1}
\end{figure}

Figure 1(a) shows the dependence of the sample resistance on gate
voltage, with a characteristic peak in the region of
electro-neutrality, where the type of carrier changes from electrons
(at positive $V_g$) to holes (at negative $V_g$). (A small offset in
the position of the maximum of $\sim -3$ V, caused by unintentional
doping of the flake \cite{NovoselovS}, is compensated in Fig. 1
(a).) The carrier mobility outside the electro-neutrality region,
$|V_g|\geq 5$ V, is $\sim$ 7000 cm$^2$V$^{-1}$s$^{-1}$. The gate
voltage dependence of the resistance in strong magnetic field has
shown quantum Hall plateaus that indicate that the sample is indeed
bilayer graphene: the conductance quantisation $\sigma_{xy}=4N
e^2/h$, where $N$ is an integer, occurs at Landau-level filling
factors $\nu=1/2, 3/2, \ldots$ \cite{NovoselovP}. An example of such
dependence with a clear $\nu=1/2$-plateau is given in Fig. 1(b).

In our study of WL we concentrate on a small magnetic field range
indicated by the box in Fig. 1(c). Because of the mesoscopic nature
of the small sample, its conductance shows reproducible fluctuations
of the order of $e^2/h$, both as a function of $V_g$ and $B$,
decaying with increasing temperature, Fig. 1(a,c).  The presence of
the fluctuations  made the magnetoconductance traces sensitive to
small shifts in $V_g$. Therefore, we have averaged the effect of
magnetic field over the range $ \Delta V_g = 2$ V, which is not too
large to change significantly the average resistance of the sample,
but large enough to contain several $R(V_g)$-fluctuations
corresponding to several ``sample realisations''. Five studied
$V_g$-regions are shown in Fig. 1(a).

\begin{figure}[htb]{}
\includegraphics[width=.9 \columnwidth]{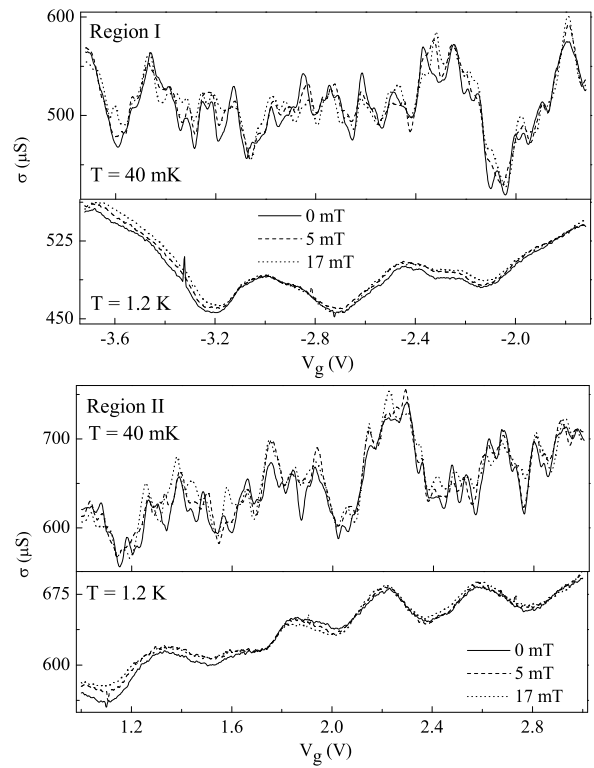}
\caption{Evolution of conductance fluctuations with small magnetic
fields applied, for regions I and II.}\label{fig:G2}
\end{figure}

Figure 2 illustrates our procedure of averaging. At a given
temperature, we find the conductivity $\sigma(V_g)$ at several
values of $B$-field, which is changed up to $\pm\, 17$ mT with a
step of $\sim 1$ mT. Then the average difference $\Delta\sigma(B)=
\langle \sigma(B,V_g)-\sigma(0,V_g) \rangle_{\Delta V_g}$ is
determined.  An example of the resulting magnetoconductivity (MC) at
several temperatures for regions I and II is presented in Fig. 3. To
decrease further the effect of conductance fluctuations, the data in
region II are averaged with those in region V which is approximately
symmetric with respect to the electro-neutrality point. (To find
$\sigma$ from the measured resistance $R(V_g)$, the resistance of
the contacts, $\sim 175$ Ohm, is taken into account. This value is
found from the deviation of the plateau in Fig. 1(b) from the
expected quantised value of 4$e^2/h$.)

\begin{figure}[htb]{}
\includegraphics[width=.9\columnwidth]{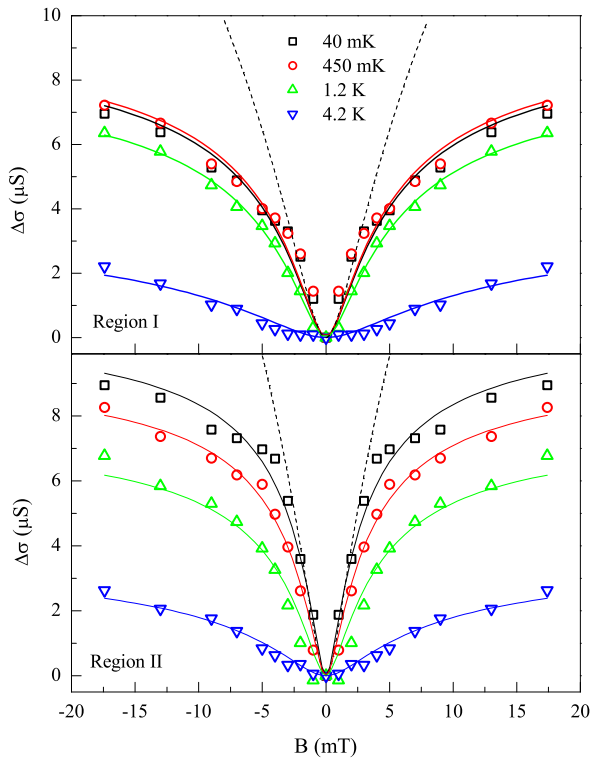}
\caption{(color online) Averaged magnetoconductivity for regions I
and II. Dashed curves are the fits using only the first term in Eq.
(1), and solid lines are the fits with the first two
terms.}\label{fig:G3}
\end{figure}

To analyse the results we use the following expression for the MC
due to WL in bilayer graphene \cite{Kechedzhi}:
\begin{eqnarray}
\Delta\sigma(B)=\frac{e^2}{\pi
h}\left[F\left(\frac{B}{B_{\phi}}\right)-F\left(\frac{B}{B_{\phi}+2B_i}\right)\right.\nonumber\\
\left.+2F\left(\frac{B}{B_{\phi}+B_{*}}\right)\right],\\
F(z)=\ln{z}+\psi{\left(\frac{1}{2}+\frac{1}{z}\right)},
\,\,B_{\phi,i,*}=\frac{\hbar}{4De}\tau^{-1}_{\phi,i,*}\nonumber.
\end{eqnarray}
Here $\psi(x)$ is the digamma function, $\tau_{\phi}$ is the phase
breaking time, $\tau_i$ is the  intervalley scattering time, and
$(\tau_*)^{-1}=(\tau_i)^{-1}+(\tau_w)^{-1}$, where $\tau_w$ is the
intra-valley ``warping'' time combined with the time of chirality
breaking \cite{Morpurgo}. Equation (1) contains three terms, and its
main difference from the single-layer result \cite{McCann} is in the
positive sign of the third term. It is seen that in the case of
$\tau_{i}, \tau_{*}\rightarrow\infty$, i.e. in the absence of the
corresponding elastic scattering, the MC is reduced to the third
term and becomes the conventional expression for WL in a 2D system
with two valleys, with a positive MC. On the contrary, strong
warping scattering ($\tau_w\rightarrow 0$) without intervalley
scattering ($\tau_i\rightarrow\infty$) suppresses the third term
while the first two cancel each other -- the result is zero MC
(complete suppression of WL). However, introducing some intervalley
scattering ($\tau_i\sim \tau_{\phi}$) will make the second term
smaller than the first, so that there will be no full cancellation
and  the first term will dominate in small fields.  The result will
be a conventional, positive MC, although with a suppression factor
of 1/2. We will now see that it is exactly this situation that
describes the experimental results.

Figure 3 shows that the MC in very small fields can be indeed
satisfactorily approximated by the first term in Eq. (1), but with
increasing $B$ this approximation strongly deviates towards larger
values. Inclusion of the second (negative) term allows one to get
good agreement with experiment over the whole field range and obtain
values of the temperature-dependent dephasing time $\tau_{\phi}$ and
temperature-independent intervalley time $\tau_i$. (Somewhat better
agreement  can be achieved by including the third term of Eq. (1),
from which the value of the warping time can be estimated -- see
below.)

\begin{figure}[htb]{}
\includegraphics[width=.98\columnwidth]{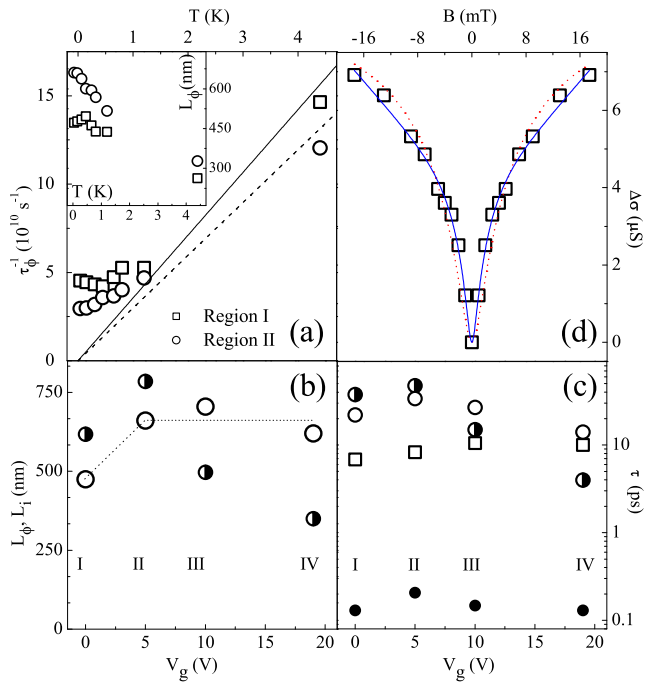}
\caption{(color online) (a) Temperature dependence of the dephasing
rate for regions I (solid line) and II (dashed line) fitted with a
linear dependence. Inset: Coherence length as a function of
temperature for the two $V_g$-regions. (b) Coherence length at 40 mK
(open circles) and intervalley scattering length (half-filled
circles) for different $V_g$-regions. (Dotted line is a guide to the
eye.) c) Dephasing time at 40 mK (open circles) and 4.2 K (squares)
in comparison with the intervalley time (half-filled circles) and
momentum relaxation time (filled circles). (d) Comparison of the
fits with two terms (dotted line) and three terms (solid line) in
Eq. (1), for the 40 mK curve in region I, Fig. 3.}\label{fig:G4}
\end{figure}

We start the discussion of the results by noting that the
conductivity of our sample is always larger than $e^2/h$, Fig. 1,
and therefore the pertubation theory of WL is applicable in the
whole range of $V_g$. Analysis of the data with the first two terms
in Eq. (1) (assuming strong warping, $\tau_w\rightarrow0$) gives the
value of the coherence length $L_{\phi}=(D \tau_{\phi})^{1/2}$ and
the phase breaking time $\tau_{\phi}$, Fig. 4(a). The diffusion
coefficient is found from the relation $D=\sigma \pi \hbar^2/2m^*
e^2$, where $\sigma$ is the average conductivity in each region and
$m^{*} = 0.033$ $m_0$ is the effective mass of carriers in graphene
bilayers \cite{McCann06Koshino}.

The manifestation of WL in region I deserves special consideration.
It is reasonable to assume that when the average carrier density is
close to zero, the presence of disorder splits the system into
electron-hole puddles \cite{Hwang} separated by p-n junctions
\cite{Cheianov,Katsnelson}. (A rough estimation of the carrier
density in the puddles, $n,p \leq 3 \times 10^{11}$cm$^{-2}$, can be
obtained from the range $|V_g| \leq4$ V where, ignoring the
fluctuations, the dependence $R(V_g)$ in Fig. 1 flattens out.) The
fact that WL in region I is not suppressed means that breaking down
the system into puddles does not prevent interference of charge
carriers on a scale larger than the puddle size. A possible reason
for this is that the p-n junctions separating the puddles are highly
transparent (at some angles of incidence) for chiral carriers
\cite{Cheianov,Katsnelson}, and therefore carrier transitions
between the puddles can occur in a coherent way.

Figure 4(a) compares $L_{\phi}(T)$ and $\tau_{\phi}(T)$ in regions I
and II. It is seen that the coherence length increases with
decreasing $T$ but approaches a finite value at $T\rightarrow 0$,
which is also seen as an offset in $\tau_{\phi}^{-1}(T)$. (Our
preliminary analysis of the $\sigma(B)$-fluctuations as universal
conductance fluctuations has given similar magnitude and behaviour
of $L_{\phi}(T)$.) There is a natural explanation that the increase
of $L_{\phi}$ at low temperatures is limited by the size of the
sample, $L = 1.5$ $\mu$m. The somewhat smaller saturation value of
$L_{\phi}$ in region I can be treated as an indication of the
formation of puddles in the electro-neutrality region.  Charge
carriers inside a puddle have to scatter many times from its
boundary before finding an optimal angle to penetrate the p-n
junction. This results in an effective decrease of their coherence
length.

The increase of the dephasing rate with temperature in all regions
agrees with the prediction of the theory of electron-electron
interaction in the diffusive regime \cite{Narozhny}, where $k_BT
\tau_p/ {\hbar} < 1$. This regime is indeed realised in our sample
where the parameter $k_BT \tau_p/ {\hbar}$ varies from 0.001 to 0.1.
We have approximated the experimental temperature dependence of the
dephasing rate at $T> 1$~K as $\tau_{\varphi}^{-1} = \beta k_BT
\ln{g} /(\hbar g)$, where $g=\sigma h/e^2$ is the dimensionless
conductivity. The empirical coefficient $\beta$ is found to be close
to unity: $\sim$1.2 in all regions, Fig. 4(a).

Figure 4(b) shows the low-$T$ saturation value of $L_{\phi}$. It is
close to half the length of the sample in all regions apart from
region I where it gets smaller, which we tentatively ascribe to the
formation of the puddles. Also shown is the temperature independent
intervalley diffusion length $L_i=(D\tau_i)^{1/2}$. This can be
interpreted as a characteristic distance between defects that are
able to change significantly the momentum of charge carriers. By
comparing this distance with the size of the sample, one can
estimate the number of such defects: $\sim10$.

Figure 4(c) compares, in several $V_g$-regions, the phase breaking
time $\tau_{\phi}$ with the intervalley scattering time $\tau_i$ and
momentum relaxation time $\tau_p$ . One can see that $\tau_{\phi}$
and $\tau_i$ are of the same order of magnitude and much larger than
$\tau_p\sim 0.15$ ps. An estimation of the warping time can be
obtained by taking into consideration the third term in Eq. (1). It
has less effect on the description of the MC and does not change
much the values of $\tau_{\phi}$ and $\tau_i$ found earlier. Figure
4 (d) illustrates the degree of improvement of the fit with
inclusion of the third term. By considering its contribution, we
obtain the following estimation of the warping time in the studied
$V_g$-regions: $\tau_*\sim\tau_w\lesssim 0.5$ ps, which is close to
the estimation \cite{Kechedzhi} of the warping time in bilayer
graphene at densities $\sim$10$^{12}$ cm$^{-2}$: $\tau_w\sim
\tau_p$. The small value of the obtained $\tau_w$ is consistent with
the smallness of the third term and is a clear signature of the
importance of the warping in the studied system.

It is interesting to compare our values of $\tau_i$ with that
recently obtained from the analysis of WL in single-layer graphene
\cite{Wu}. The sample in \cite{Wu} was fabricated by a different
method (thermal decomposition of SiC)  and studied at a larger
electron density, $n =3.4 \times 10^{12}$ cm$^{-2}$. The dephasing
time (for similar temperatures) and the momentum relaxation time are
close in both cases. The intervalley scattering time, however, in
our experiment is 10-30 times larger than in \cite{Wu}. As the
intervalley scattering is controlled by atomically sharp impurities,
the difference in the concentration of these impurities can be due
to the difference in the fabrication techniques. Alternatively, the
difference in $\tau_i$ can be caused by the intrinsic difference
between single-layer and bilayer: scattering by sharp defects can be
of less importance in a more robust system with two layers.

In conclusion, we have performed an experimental investigation of
the magnetoresistance caused by weak localisation in bilayer
graphene controlled by gate voltage. Several ranges of $V_g$,
including the electro-neutrality region, have been studied. Firstly,
we have found that WL is not suppressed in all regions. Secondly, we
have found good agreement of our results with the prediction of the
theory of WL in bilayer graphene \cite{Kechedzhi}, and have
determined the values of the characteristic times which make the
manifestation of WL in bilayer graphene different from that in
conventional 2D systems.

We are grateful to A. Geim, K. Novoselov and D. Jiang for sharing
their knowledge of  the fabrication of graphene flakes by mechanical
exfoliation. We are indebted to V. I. Fal'ko, E. McCann and K.
Kechedzhi for discussing their results prior to publication. We also
acknowledge useful discussions with V. V. Cheianov, P. Kim, F.
Guinea, A. Castro Neto concerning the physics of graphene-based
structures.

 \newpage

\end{document}